\documentclass[%
reprint,
 amsmath,amssymb,
 aps,
]{revtex4-1}

\usepackage{graphicx}% Include figure files
\usepackage{dcolumn}% Align table columns on decimal point
\usepackage{bm}% bold math

\def\beq{\begin{equation}}
\def\eeq{\end{equation}}
\DeclareMathOperator{\hf}{\frac{1}{2}}
\DeclareMathOperator{\sd }{\partial_\Lambda }
\DeclareMathOperator{\Rd}{\partial^R_\Lambda}
\DeclareMathOperator{\Tr}{\text{Tr}}
\DeclareMathOperator{\STr}{\text{STr}}
\DeclareMathOperator{\Ax}{\mathcal{A}_\textit{x}}
\DeclareMathOperator{\At}{\mathcal{A}_\tau}
\DeclareMathOperator{\Bx}{\mathcal{B}_{\textit{x}}}
\DeclareMathOperator{\Bt}{\mathcal{B}_\tau}
\DeclareMathOperator{\nft}{\eta_{\tau}^{\textit{f}}}
\DeclareMathOperator{\nfx}{\eta_{\textit{x}}^{\textit{f}}}
\DeclareMathOperator{\nbt}{\eta_{\tau}^{\textit{b}}}
\DeclareMathOperator{\nbx}{\eta_{\textit{x}}^{\textit{b}}}

\newcommand{\tcite}[1]{~\cite{#1}}

\newcommand{\pder}[1]{\frac{\partial}{\partial #1}}

\begin{document}

\title{Non-Fermi liquid fixed points and anomalous Landau damping in a quantum critical metal}% Force line breaks with \\

\author{Matthew J. Trott}
\affiliation{SUPA, School of Physics and Astronomy, University of St Andrews, North Haugh, St Andrews, Fife KY16 9SS, United Kingdom}
\author{Chris A. Hooley}
\affiliation{SUPA, School of Physics and Astronomy, University of St Andrews, North Haugh, St Andrews, Fife KY16 9SS, United Kingdom}

\date{Thursday 1st November 2018}% It is always \today, today,

\begin{abstract}
	\noindent
	We present a functional renormalization group calculation of the properties of a quantum critical metal in $d=2$ spatial dimensions. Our theory describes a general class of Pomeranchuk instabilities with $N_b$ flavors of boson. At small $N_b$ we find a family of fixed points characterized by weakly non-Fermi-liquid behavior of the conduction electrons and $z \approx 2$ critical dynamics for the order parameter fluctuations, in agreement with the scaling observed by Schattner {\it et al.}\ [Phys.\ Rev.\ X {\bf 6}, 031028 (2016)] for the Ising-nematic transition. Contrary to recent suggestions that this represents an intermediate regime en route to the scaling limit, our calculation suggests that this behavior may persist all the way to the critical point.  As the number of bosons $N_b$ is increased, the model's fixed-point properties cross over to $z \approx 1$ scaling and non-Fermi-liquid behavior similar to that obtained by Fitzpatrick {\it et al.\/}\ [Phys. Rev. B \textbf{88}, 125116 (2013)].
\end{abstract}

\maketitle

\noindent
{\it Introduction.} A major question in modern condensed matter physics is how to describe states of matter that are metallic but do not have coherent fermionic quasiparticle excitations.  These `non-Fermi liquids' appear in several families of materials, including the cuprates\tcite{Sachdev2010}, the heavy-fermion compounds\tcite{Gegenwart2008}, and the iron-based superconductors\tcite{Shibauchi2013}. They are also important in nuclear physics, where they describe dense quark matter interacting with gauge fields\tcite{Schafer2014}.

A common approach to formulating theories of these non-Fermi-liquid states is to regard them as arising from the interaction of the electronic excitations of a conventional metal with the slow bosonic fluctuations associated with incipient long-range order (ferromagnetic, antiferromagnetic, nematic, etc.,\ depending on the material in question).  In such approaches, fermion-fermion interactions are assumed to be mediated only via these bosonic modes, which may be thought of as resulting from a Hubbard-Stratonovich decoupling of the original four-fermion interactions.

The modern formulation of Landau's Fermi-liquid theory is in terms of a renormalization group (RG) fixed point for which all interactions are irrelevant, except for a low-temperature instability to superconductivity\tcite{Shankar1991,Polchinski1992,Shankar1994}. Close to a quantum critical point (QCP), however, additional degrees of freedom arise, causing the breakdown of Landau's theory and the occurrence of non-Fermi-liquid physics. 

Traditionally the analysis of such mixed fermion-boson theories is performed by integrating out the fermionic degrees of freedom to produce an effective order-parameter theory of the quantum critical metal\tcite{Hertz1976,Millis1993}.  The bosonic propagator gains a Landau-damping term with a dynamical exponent $z$ which encodes the decay of the order parameter into particle-hole excitations.  However, integrating out gapless modes on the Fermi surface causes non-analytic and singular corrections to the effective order parameter\tcite{Belitz1997,Abanov2004,Thier2011}. 

Because of these issues, interest has recently grown in approaches that retain the fermionic degrees of freedom and treat the fermions and bosons on an equal footing.  A particular case of interest is that of a metal approaching the transition to Ising-nematic order.  In two spatial dimensions the electron nematic was first predicted to exhibit overdamped $z=3$ dynamics of the boson with a fermionic self-energy of the form $\omega^{2/3}$\tcite{Historical}. However, within the field-theoretic RG treatment it was then discovered that high-loop diagrams contribute even when the number of fermion flavors is large\tcite{Lee2009}, and singular corrections arise in the fermion and boson self-energies\tcite{Metlitski2010,Holder2015}, raising doubts about the validity of these conclusions.  Further, recent quantum Monte Carlo studies have suggested the theory is governed by $z=2$ critical dynamics\tcite{Schattner2016}. $z=2$ dynamics has so far only been found in theories with many boson flavors\tcite{Fitzpatrick2014}; however, as we show in this paper, this may not be the whole story.

The status of Wilsonian approaches to the problem appears, if anything, even less clear.  Wilsonian effective field theories cannot develop non-local or non-analytic corrections, and it is not obvious how $z=3$ boson dynamics could arise during the flow. Attempts to combine Wilsonian and perturbative methods\tcite{Ridgway2015} have provided novel results but have not addressed the nature of how $z\neq1$ dynamics can arise from the local ultra-violet (UV) theory. Previous perturbative Wilsonian analysis starting with local propagators\tcite{Fitzpatrick2013} found no departure from $z=1$ dynamics:\ the infra-red (IR) fixed point was found to be of the Wilson-Fisher type, with a fermionic self-energy of the form $\omega^{3/4}$, which disagrees with the result of the field-theoretic RG. 

\begin{figure}[t]
	\begin{center}
		\includegraphics[width=0.85\columnwidth]{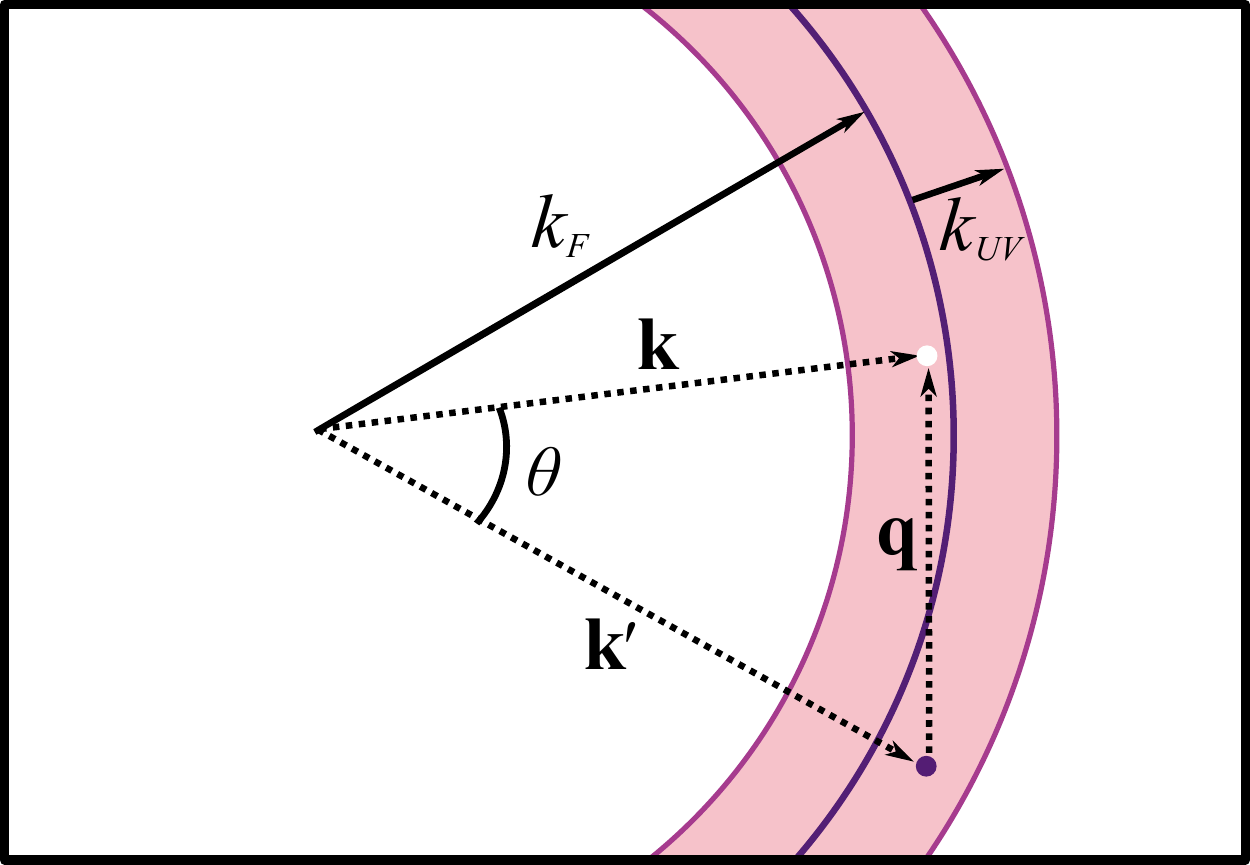}
		\caption{A portion of the Fermi surface of our model, showing the two relevant high-energy scales in the problem:\ $k_F$, the Fermi momentum, and $k_{\rm UV}$, the upper limit of validity of our starting Fermi-liquid theory.  These enter our fixed-point results as a dimensionless ratio, $N \equiv k_F/k_{\rm UV}$.  The figure also shows an example of particle-hole annihilation by a boson with momentum $\mathbf{q}=\mathbf{k}-\mathbf{k'}$.}
		\label{fermisurface}
	\end{center}
\end{figure}

\begin{figure}[t]
	\centering
	\includegraphics[width=0.48\columnwidth]{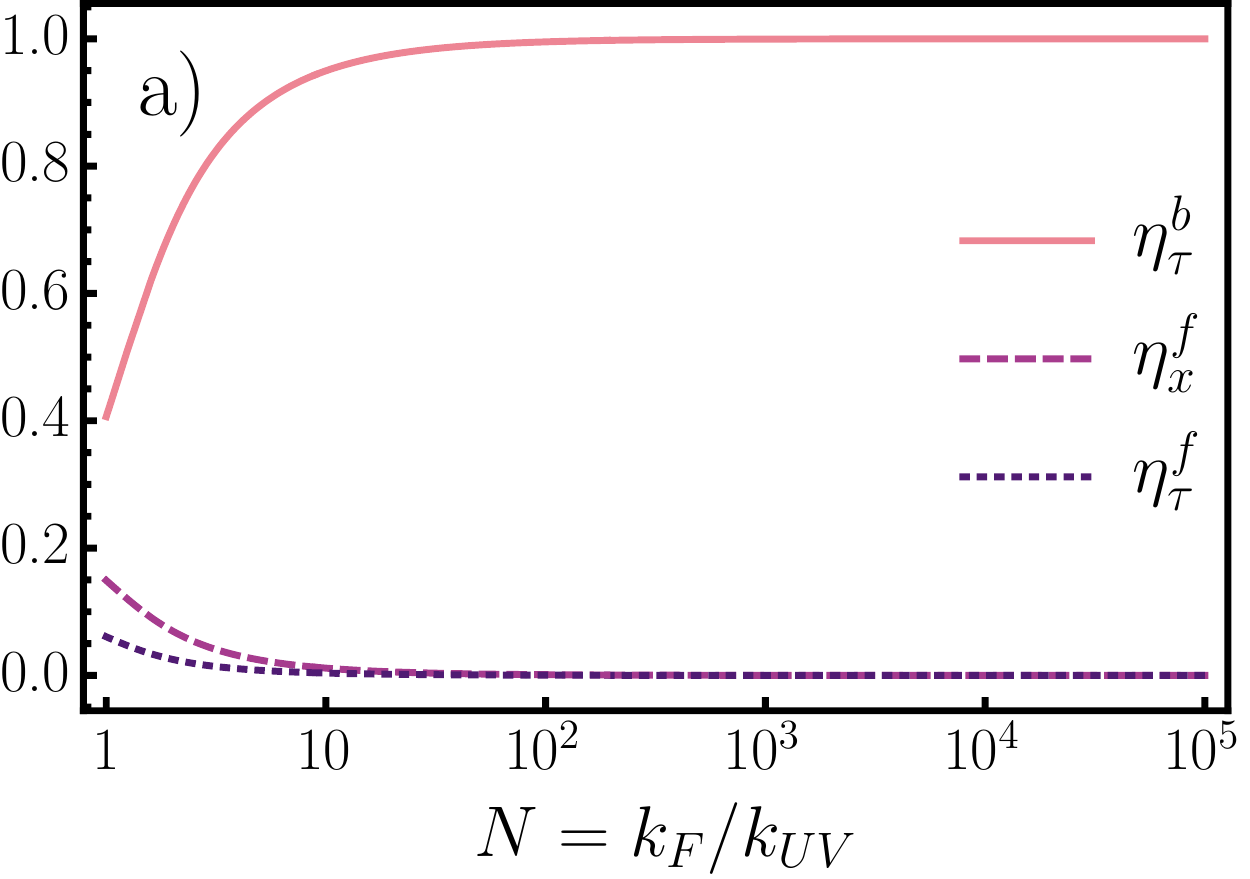}\quad
	\includegraphics[width=0.48\columnwidth]{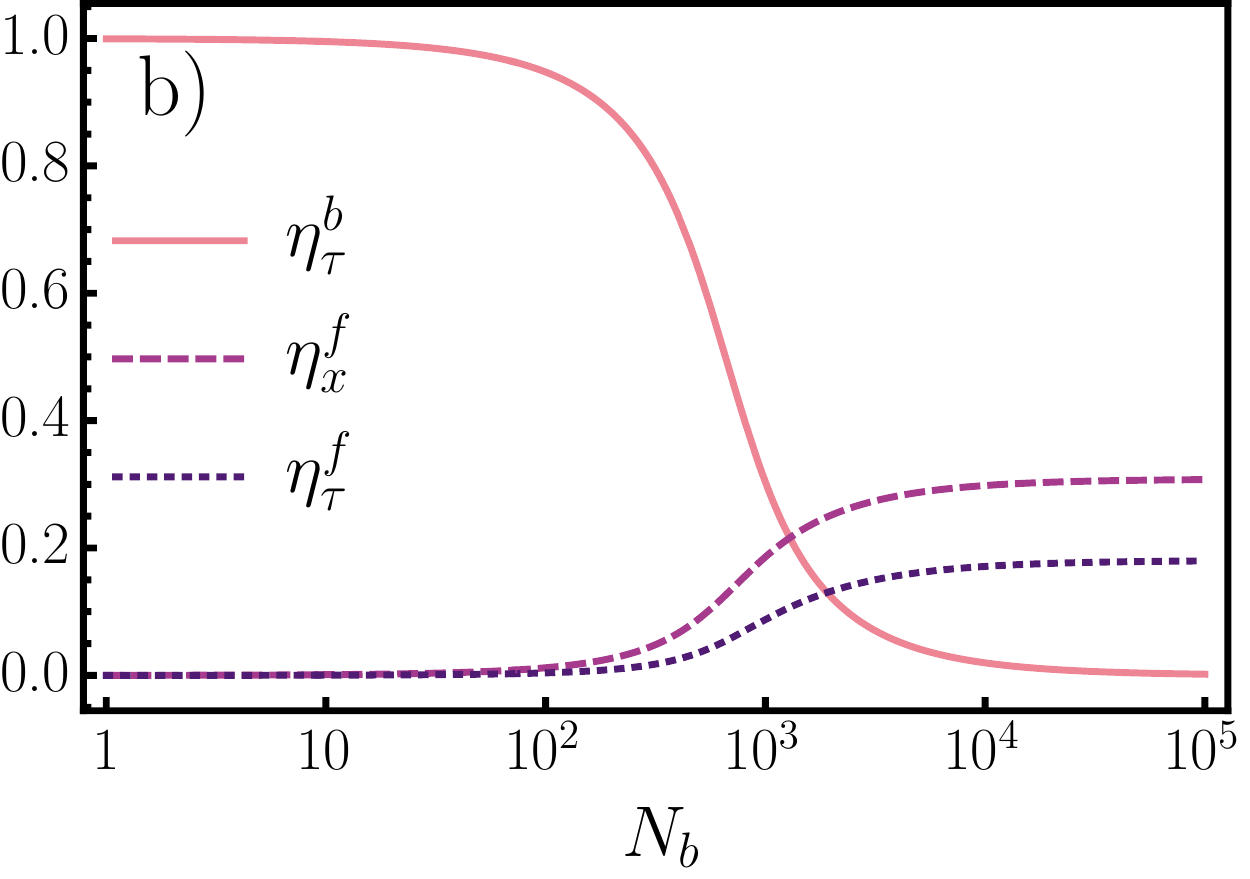}\vspace{0.2em}
	\includegraphics[width=0.475\columnwidth]{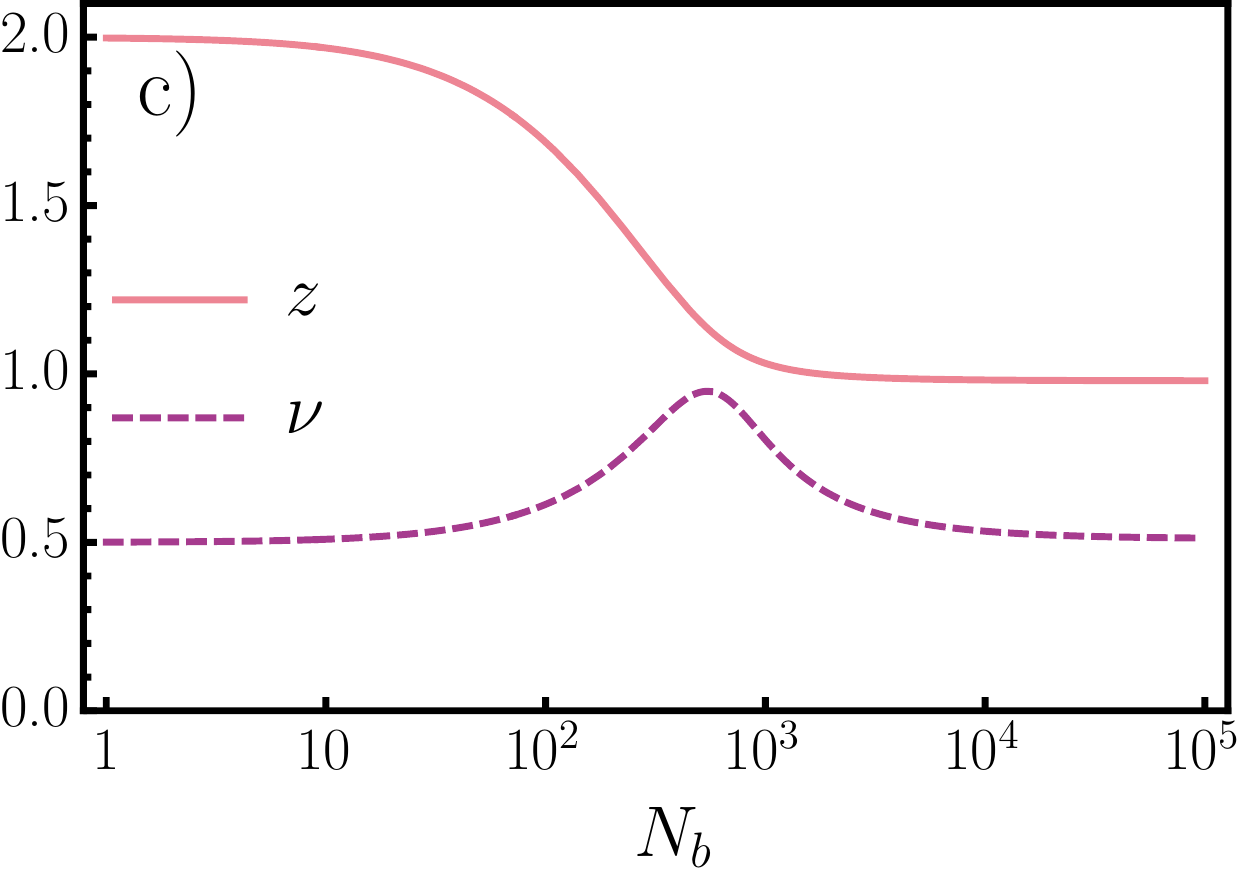}\quad
	\includegraphics[width=0.485\columnwidth]{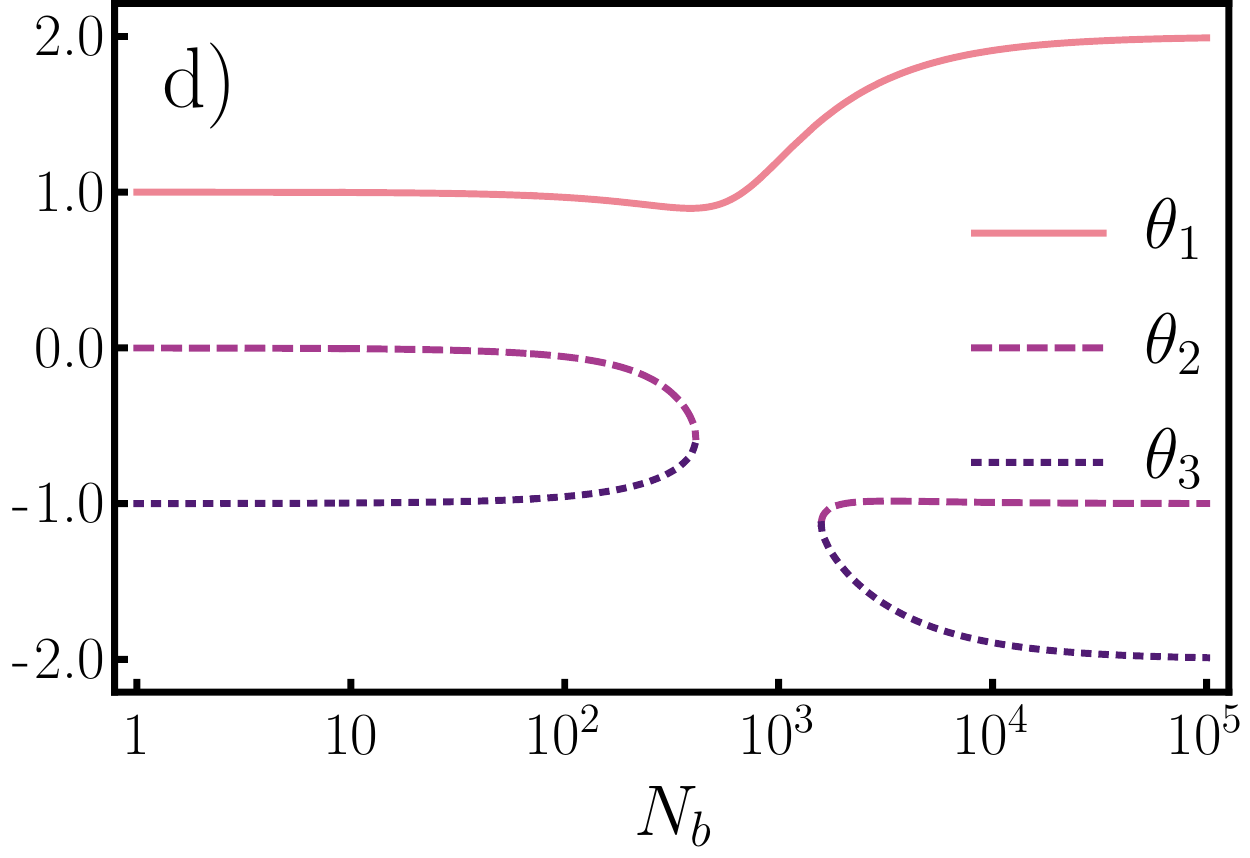}
	\caption{(a) The anomalous dimensions $\eta^b_\tau$, $\eta^f_x$, and $\eta^f_\tau$ as functions of $N \equiv k_F/k_{\rm UV}$ for the case of a single flavor of boson, $N_b=1$.  (b) The same anomalous dimensions as functions of $N_b$ for $N = 10^3$, showing the crossover from weak to strong non-Fermi-liquid behavior at $N_b \sim N$. (c) The dynamical exponent $z$ and the critical exponent $\nu$ as functions of $N_b$ for $N = 10^3$, showing the crossover from $z=2$ to $z=1$ boson dynamics at $N_b \sim N$. (d) The eigenvalues of the stability matrix as functions of $N_b$ for $N=10^3$.}
	\label{anomdim}
\end{figure}

In this work we demonstrate the links between some of the above results, via a functional RG (fRG) analysis of a quantum critical metal with $N_b$ bosonic flavors in $d=2$ spatial dimensions.  As well as $N_b$, our results also depend on another parameter, $N \equiv k_F/k_{\rm UV}$, where $k_F$ is the radius of the Fermi surface and $k_{\rm UV}$ is the momentum scale characterizing the limit of validity of our starting Fermi-liquid theory (see Fig.~\ref{fermisurface}). For $N_b \ll N$, we find behavior reminiscent of the numerical results of\tcite{Schattner2016}:\ weakly non-Fermi-liquid conduction electrons and $z \approx 2$ critical dynamics for the bosonic order parameter fluctuations.  For $N_b \gg N$, we approach $z \approx 1$ scaling and non-Fermi-liquid behavior similar to\tcite{Fitzpatrick2014}.  Importantly, our theory is able to describe the crossover between these two types of behavior as $N/N_b$ decreases (see Fig.~\ref{anomdim}).

{\it Model and methods.}
Our effective field theory describes a Pomeranchuk instability, i.e.\ the breaking of Fermi-surface rotational symmetry. Pomeranchuk instabilities are classified by their angular momentum channel $l$; the Ising-nematic transition, for example, is the $N_b=1$, $l=2$ case of our model. As the momentum dependence of the coupling $g(k,q)$ becomes irrelevant at low energies, the results obtained for the $l=0$ instability $g(k,q)\approx g$ become general for all channels\tcite{Fitzpatrick2013}.  In many aspects our calculation follows that of Maier and Strack\tcite{Maier2016}, who performed a similar analysis for a cuprate-like strange metal with a spin-density-wave instability. Some of the more technical details of our calculation may be found in the Supplemental Material.

We assume a circular Fermi surface\tcite{Polchinski1992}; the entire Fermi surface is retained, and thus additional constraints on scaling imposed in patch-scheme studies are avoided.  We linearize the dispersion near the Fermi surface:\ $\xi_{\mathbf{k}}\approx\Ax\ell$ with $\mathbf{k}=\mathbf{\hat{n}}(k_F+\ell)$, such that $\ell$ is a perpendicular displacement in the direction $\mathbf{\hat{n}}$ from the Fermi surface to $\mathbf{k}$. $\ell$ scales towards the Fermi surface under the RG. The Fermi momentum $k_F$ is fixed and does not scale. Momenta parallel to the Fermi surface are instead parameterized by angles as is usual in Fermi liquid theory.

The linearization procedure is valid for $|\ell|<k_{\rm UV}$, where $k_{\rm UV}$ is the momentum scale above which the Fermi-liquid effective field theory breaks down\tcite{Kopietz1997}. Thus $k_{\rm UV}$ and $k_F$ replace the microscopic degrees of freedom of the system with parameters that can ideally be determined from experiment. The low-energy theory is thus intrinsically $k_{\rm UV}$- and $k_F$-dependent, and the fixed-point properties we obtain are dependent on the non-universal quantity $N=k_F/k_{\rm UV}$. The non-universality arises due to a combination of UV/IR mixing in schemes with Landau damping\tcite{Fitzpatrick2015} and the choice of a frequency cutoff. Previous momentum-cutoff schemes have found a dependence of the low-energy parameters on $k_F$\tcite{Mandal2015}. The frequency-cutoff scheme is instead dependent on the dimensionless ratio $N$. The non-universality also appears in the RG of Shankar\tcite{Shankar1994} except with $k_{\rm UV}$ as the flow parameter. This allows for a $1/N$ expansion in the low-energy limit as $k_{\rm UV}$ is lowered, suppressing large classes of diagrams. Additionally the purely fermionic theory allows for a rescaling of parameters to eliminate $k_F$. The addition of bosonic degrees of freedom removes this possibility.  

We use the fRG formulated by Wetterich\tcite{Wetterich1993}, which is a modern implementation of Wilsonian renormalization which iteratively integrates out degrees of freedom within the functional-integral representation of the theory's partition function. The governing flow equation,
\begin{equation}\label{flow}
	\sd  \Gamma_\Lambda=\hf\STr\left(\Gamma_\Lambda^{(2)}+R_\Lambda\right)^{-1}\sd   R_\Lambda,
\end{equation}
takes the form of a functional differential equation for the effective action $\Gamma$, the generator of one-particle-irreducible correlation functions.  Here the supertrace is defined by $\STr A=\Tr A_b-\Tr A_f$, with $A_b$ and $A_f$ the bosonic and fermionic sectors of the matrix $A$.  The trace $\Tr$ denotes a sum over field degrees of freedom and integration over frequencies and momenta. The Hessian $\Gamma_\Lambda^{(2)}=\frac{\overrightarrow{\delta}}{\delta \bar{\chi}}\Gamma_\Lambda\frac{\overleftarrow{\delta}}{\delta \chi}$ is a matrix of functional derivatives with respect to superfields $\chi$ and $\overline{\chi}$ composed of the fermionic and bosonic degrees of freedom.

The effective average action $\Gamma_\Lambda$ flows from the microscopic action $\Gamma_{\Lambda\rightarrow\Lambda_{\rm UV}}=S$ at UV scale $\Lambda_{\rm UV}$ to the quantum effective action $\Gamma_{\Lambda\rightarrow0}=\Gamma$ in the IR. The regulator function $R_\Lambda$ is introduced to induce the flow of $\Gamma_\Lambda$ and cuts off IR divergences, suppressing frequencies $|\omega|<\Lambda$\tcite{Metzner2012}. 

The form of the regulator function $R_\Lambda$ is an important feature of our RG scheme.  In non-relativistic fRG schemes, momentum cutoffs have been found to suppress the fermionic soft modes that give rise to the Landau-damping of the order parameter fluctuations\tcite{Honerkamp2001,Husemann2009}.  This is because the low-energy behavior of the order parameter is governed by a single point in momentum space, whereas fermionic properties are determined by gapless excitations along the entire Fermi surface, a manifold with codimension one, a discrepancy that is hard to deal with in momentum-based RG schemes.  For these reasons, and following Maier and Strack\tcite{Maier2016}, we use a frequency cutoff, which allows us to capture the soft-mode excitations as they become singular towards $\mathbf{q}={\bf 0}$.

Our fermion regulator function takes the following form:
\beq\label{regf}
R_f=[i\At \omega-\Ax\ell]\left[\chi^{-1}(\omega,\Lambda)-1\right],
\eeq 
with 
\beq 
\chi(\omega,\Lambda)=\frac{\omega^2}{\omega^2+\Lambda^2}.
\eeq
Both bosonic and fermionic frequencies scale towards zero, meaning that the complexity of the scaling to the Fermi surface manifold is avoided. Our bosonic regulator function is $R_b=\Bt\Lambda^2$.

The action for the quantum critical metal is $\Gamma_\Lambda=\Gamma_f+\Gamma_b+\Gamma_g+\Gamma_{\lambda}$, where
\begin{align}
	\Gamma_f &=\sum_{\mu=\uparrow\downarrow}\int_k\overline{\psi}_{k,\mu}\left[i\At \omega-\Ax\ell\right]\psi_{k,\mu},\\
	\Gamma_b &=\hf\sum_a\int_q\phi^a_q\left[\Bt\Omega^2+\Bx q^2+\tilde{\delta}\right]\phi^a_{-q},\\
	\Gamma_g &=\sum_{a,\mu}\int_{k,q}\tilde{g}(k,q)\phi^a_q\overline{\psi}_{k+q,\mu}\psi_{k,\mu},\\
	\Gamma_{\lambda} &=\tilde{\lambda}\sum_{a,b}\int_{q_1,q_2,q_3}\phi^a_{q_1+q_3}\phi^a_{q_2-q_3}\phi^b_{q_1}\phi^b_{q_2}. 
\end{align} 
Thus the action $\Gamma$ is parameterized by seven renormalization constants which depend on the running scale $\Lambda$. The fermion and boson fields are coupled by a local Yukawa interaction, and we have truncated the bosonic self-interactions at quartic order.

The parameters $\At$ and $\Ax$ renormalize the frequency and momentum dependence of the fermion propagator independently. They can be expressed in Fermi liquid form with quasiparticle weight $Z=1/\At$ and Fermi velocity $v=\Ax/\At$. The $\Lambda$ dependence of the parameters has been suppressed for brevity.

The bosonic propagator is parameterized by three scale dependent factors $\Bt$, $\Bx$ and $\tilde{\delta}$. The order parameter is an $N_b$-component symmetric scalar field with velocity $c^2=\Bx/\Bt$ describing collective excitations in the symmetric phase. No gapless fermionic modes have been integrated out so the propagator is fully local and has the dynamical exponent $z=1$. This correctly describes the physics in the high-energy limit of the theory. The frequency and momentum terms are again allowed to renormalize separately. The dimensional mass $\tilde{\delta}$ vanishes as $\Lambda\rightarrow0$ to reach criticality.

Additional four-point fermion vertices generated in the flow are neglected.  This allows us to retain a minimal model for the electron-boson system describing only the critical state at the transition, but means that we ignore eventual pairing instabilities that may set in close to the critical point.

Imposing that momenta and frequency scale in the same way, as $[\omega]=[\Omega]=[\ell]=[|\mathbf{q}|]=1$, the fields and boson mass scaling dimensions in terms of $\Lambda$ are
\begin{align}
	[\psi_{k,\mu}]=-\frac{3}{2},&& [\phi^a_q]=-\frac{5}{2},&& [\tilde{\delta}]=2.
\end{align}
For large momentum transfers parallel to the Fermi surface the Yukawa coupling is irrelevant\tcite{Polchinski1992,Fitzpatrick2013}. For small momentum transfers with $\mathbf{\hat{n}}\approx\mathbf{\hat{n}'}$ the full momentum dependence of the Yukawa coupling $\tilde{g}(k,q)=\tilde{g}+\tilde{a}_1\ell+\tilde{a}_2|\mathbf{q}|+\dots$ can be ignored as only the constant part $g$ is relevant and all higher-momentum terms in the expansion are irrelevant. 
\begin{align}
	[\tilde{g}]=\hf,&& [\tilde{a}_1]=-\hf,&& [\tilde{a}_2]=-\hf.
\end{align}
The scaling dimension of the boson self interaction is $[\tilde{\lambda}]=1$.

The beta functions and anomalous dimensions can be calculated analytically in the IR asymptotic limit and for ratio $c^2/v^2=\Bx\At^2/\Bt\Ax^2\rightarrow0$. Taking this limit is consistent with the overdamped boson dynamics seen in previous theoretical studies\tcite{Historical,Metlitski2010}. $c^2/v^2$ is dependent on the scale $\Lambda$ with the power $[c^2/v^2]=\nbt+2\nfx-\nbx-2\nft$. Figs.~\ref{anomdim}(a) and \ref{anomdim}(b) show that the vanishing of this ratio in the IR is self-consistent for $1\leqslant N\leqslant10^5$ and $1\leqslant N_b\leqslant10^5$.

{\it Results.}
The beta functions $\Lambda\sd\lambda_{i}=\beta_{i}$ for the dimensionless parameters 
\begin{align}\label{rescaled}
	\delta=\frac{\tilde{\delta}}{\Bt\Lambda^2},&& \lambda=\frac{\tilde{\lambda}}{2\pi\Bx\Bt\Lambda},&&  g^2=\frac{N\tilde{g}^2}{8\pi\Ax^2\Bt\Lambda}
\end{align}
are given by
\begin{align}\label{betad}
	\beta_{\delta }&=(-2+\nbt)\delta -(2+N_b) (2-\nbt)\frac{\lambda}{\sqrt{1+\delta }}+12g^2,\\ \label{betal}
	\beta_{\lambda }&=(-1+\nbt)\lambda +(8+N_b)(2-\nbt)\frac{\lambda ^2}{2(1+\delta )^{3/2}},\\ \label{betag}
	\beta_{g^2}&=\left(-1+\nbt+2\nfx-2\eta_g\right)g^2,
\end{align}
with
\beq \label{ng}
\eta_g=-\frac{N_bg^2}{ N}\frac{ \left(3 \sqrt{1+\delta}+3-\nbt\right)}{  \left(\sqrt{1+\delta}+1\right)^3}.
\eeq

The boson anomalous dimensions, defined as $\nbt=-\Lambda\sd\ln\Bt$, $\nbx=-\Lambda\sd\ln\Bx$, are given by 
\begin{align}\label{nb}
	\nbt=g^2,&& \nbx=0.
\end{align}

For the fermionic sector, the momentum anomalous dimension $\nfx=-\Lambda\sd\ln\Ax$ is given by
\beq \label{nfx}
\nfx=\frac{N_bg^2}{N}\frac{ \left(2\sqrt{1+\delta}+2-\nbt\right)}{ \sqrt{1+\delta} \left(\sqrt{1+\delta}+1\right)^2}.
\eeq
The frequency anomalous dimension $\nft=-\Lambda\sd\ln\At$ is given by
\beq \label{nft}
\nft=\frac{N_bg^2}{N}\frac{  \left(2 \left(\sqrt{\delta +1}+1\right)+\left(\sqrt{\delta +1}-1\right) \nbt\right)}{
	\sqrt{\delta +1} \left(\sqrt{\delta +1}+1\right)^3}.
\eeq 
The feedback of fermion anomalous dimensions $\nfx$ and $\nft$ has been ignored within the beta functions and anomalous dimensions (\ref{betad}-\ref{nft}), allowing for analytic expressions for the plots in Fig.~\ref{anomdim}. The error from discarding the feedback is exactly zero in the $N/N_b\rightarrow\infty$ limit and approximately $10\%$ in the $N/N_b\rightarrow0$ limit. The full expressions are presented in the Supplemental Material. $\nbx=0$ for all $N$ and $N_b$. The anomalous dimensions $\nbt$, $\nfx$ and $\nft$ are plotted in Fig.~\ref{anomdim}(a) for $N_b=1$, $1\leqslant N\leqslant10^5$ and in Fig.~\ref{anomdim}(b) for $N=10^3$, $1\leqslant N_b\leqslant10^5$. 

The critical exponents can be calculated from Eqs. (\ref{betad})-(\ref{betag}) via the stability matrix $\mathcal{M}_{ij}\equiv\partial\beta_{i}/\partial\lambda_j|_{\lambda=\lambda_*}$. The eigenvalues $\theta_i$ of $-\mathcal{M}_{ij}$ characterize the scaling laws at the fixed point. Positive eigenvalues correspond to relevant RG directions. As a consequence of using a frequency scale as a flow parameter the largest eigenvalue $\theta_1$ gives the exponent $\theta_1=1/\nu z$ for the behavior of the correlation time $\tau$. Were we instead using a momentum scale $\Lambda_k$ as the flow parameter, the largest eigenvalue of the $\Lambda_k$-flow stability matrix would give $1/\nu$:\ we thus see that the mapping between frequency and momentum stability matrices is non-trivial for $z\neq1$.  The critical exponent $\gamma$ is calculated for the susceptibility $\chi= 1/\tilde{\delta}$; the critical exponent $\eta$ is just equal to $\nbx$. 

At the QCP the bosonic self-interaction remains finite. This leads to the fixed point solution for the mass
\beq 
\delta_*=2\frac{2+N_b-g_*^2(50+7N_b)}{3g_*^2(4+N_b)-4(5+N_b)}.
\eeq
For $N_b=1$ and in the $N\rightarrow\infty$ limit $\delta_*\rightarrow 12$ and $g_*^2\rightarrow1$ at the fixed point, while the anomalous dimensions approach
\begin{align}
	\nbt\rightarrow1, &&\nbx=0, &&\nfx\rightarrow0, &&\nft\rightarrow0.
\end{align}
The largest eigenvalue of the stability matrix corresponding to the mass term approaches $\theta_1\rightarrow1$, and thus $\nu z\rightarrow1$. This gives the critical exponents
\begin{align}
	z=\frac{2-\nbx}{2-\nbt}\rightarrow2,&& \nu\rightarrow\hf,&& \gamma\rightarrow1,&& \eta=0.
\end{align}
The static exponents are mean-field-like, while the dynamical exponent departs substantially from the high-energy form.

Significantly, the exponents match those seen numerically by Schattner {\it et al.\/}\tcite{Schattner2016}.  They argued that their $z \approx 2$ scaling was a property of an intermediate finite-temperature regime preceding a scaling regime that was inaccessible to their quantum Monte Carlo analysis at the system sizes they could reach.  However, we find a stable QCP at $N_b=1$, which suggests that the $z=2$ regime can exist at zero temperature and all the way to the QCP.

In the $N/N_b \rightarrow0$ limit the effect of fermionic fluctuations is decreased  such that $\nbt\rightarrow0$ and a non-Fermi liquid state is formed
\begin{align}
	\nfx\rightarrow\frac{1}{17}(3\sqrt{2}+1), &&\nft\rightarrow\frac{1}{17}(5\sqrt{2}-4).
\end{align}
The boson dynamics become $z\rightarrow1$ and the fixed point values of $\delta$ and $\lambda$ become those of the decoupled Wilson-Fisher fixed point.  The crossover between these two regimes is shown in Fig.~\ref{anomdim}(c).

The eigenvalue $\theta_2$ corresponding to the bosonic interaction term is irrelevant. It becomes exactly marginal in the $N/N_b \rightarrow\infty$ limit when the interacting fixed point collides with an unstable multicritical point with $\lambda_*=0$. As $N_b$ becomes large the eigenvalues become more recognizable as those of a Wilson-Fisher fixed point, altered slightly due to the coupling to fermions. The eigenvalues are plotted as functions of $N_b$ in Fig.~\ref{anomdim}(d), for a fixed value of $N$.

\textit{Summary and discussion.} We have presented a functional RG analysis of a quantum critical metal in the vicinity of a Pomeranchuk instability.  Our results depend on the ratio between the fermionic parameter $N$ and the number of flavors of boson $N_b$. 

As $N/N_b \to \infty$, the system is described by $z=2$ boson dynamics and the conduction electrons show Fermi-liquid behavior. As $N/N_b$ is reduced, the dynamical exponent decreases and the conduction electrons become non-Fermi-liquid.  For $N/N_b \to 0$, the boson dynamics become undamped with $z=1$, and we find a non-Fermi liquid with a fermion self-energy of the form $\omega^{1-\nft}\approx\omega^{0.819}$.  This is smaller than the perturbative results obtained by Fitzpatrick {\it et al.\/}\tcite{Fitzpatrick2013}. However this is expected due to the increased effect of particle-hole fluctuations in the soft frequency scheme. 

The soft frequency regulators that we use in our calculation capture the feedback of soft particle-hole excitations on the boson, which are beyond the scope of the hard cut-off theory extrapolated to $\epsilon=1$\tcite{Fitzpatrick2013}. Therefore in our calculation the fermions renormalize the bosonic sector much more strongly, causing a departure from $z=1$ and a significant weakening of the non-Fermi liquid effects.  As the number of boson flavors is increased, the fermionic fluctuations are subdued and the theory reverts to behavior similar to the hard cut-off results.

A worthwhile extension to the work presented in this paper would be to consider the two-channel problem, retaining both forward-scattering and pairing channels. Experimentally the nematic phase has been observed in close vicinity to a superconducting state, and theory suggests that nematic fluctuations enhance superconductivity\tcite{Lederer2015}. It would therefore be interesting to investigate the interplay of nematic and superconducting phases and determine whether the Pomeranchuk instability fixed point remains stable.

\textit{Acknowledgments.} MJT acknowledges financial support from the CM-CDT under EPSRC (UK) grant number EP/L015110/1.  CAH acknowledges financial support from the TOPNES programme under EPSRC (UK) grant number EP/I031014/1.

\begin{widetext}
\newpage
\appendix
\begin{center}
	\Large M.J.\ Trott and C.A.\ Hooley,\\
	\Large ``Non-Fermi liquid fixed points and anomalous Landau damping in a quantum critical metal''\\
	\footnotesize \ \\
	\large Supplemental Material
\end{center}% Force line breaks with \\

%\maketitle

\vspace{1cm}
Equations in the main text are referred to as (M1) etc..

\section*{Computing the flow equations}

The flow equations can be computed via expanding the flow equation for the effective average action (M1)
\beq \label{expanded}
\sd\Gamma_\Lambda=\hf\Rd\STr\ln(\Gamma_{\Lambda,0}^{(2)}+\mathbf{R}_\Lambda)+\hf\Rd\STr\sum_{n=1}^{\infty}\frac{(-1)^{n+1}}{n!}\left(\frac{\Delta\Gamma_{\Lambda}^{(2)}}{\Gamma_{\Lambda,0}^{(2)}+\mathbf{R}_\Lambda}\right)^n
\eeq
where $\Gamma_{\Lambda,0}^{(2)}$ denotes the field independent propagator terms and $\Delta\Gamma_{\Lambda}^{(2)}$ denote the field dependent fluctuation terms. The regulator derivative is defined $\Rd=\sum_{i=f,b}\sd R_i\frac{\partial}{\partial R_i}$ with $R_f$ and $R_b$ defined in the main paper. The ansatz for the effective action can be plugged in to (\ref{expanded}) and the beta functions obtained by matching coefficients.

The fermionic propagator is given by
\beq 
G_f(\omega,\mathbf{k})=\frac{\chi(\omega,\Lambda)}{i\At\omega-\Ax\ell}.
\eeq 
The fermionic single scale propagator is given by
\beq 
S_f(\omega,\mathbf{k})=\Rd G_f(\omega,\mathbf{k})=\chi'(\omega,\Lambda)\frac{i(2-\nft)\At\omega-(2-\nfx)\Ax\ell}{2(i\At\omega-\Ax\ell)^2}.
\eeq
The prime denotes a scale derivative acting only on $\chi(\omega,\Lambda)=\omega^2/(\omega^2+\Lambda^2)$.

The bosonic propagator is given by
\beq 
G_b(\Omega,\mathbf{q})=\frac{-1}{\Bt(\Omega^2+\Lambda^2)+\Bx\mathbf{q}^2+\tilde{\delta}}.
\eeq 
The bosonic single scale propagator is 
\beq 
S_b(\Omega,\mathbf{q})=\Rd G_b(\Omega,\mathbf{q})=\frac{\Lambda\Bt(2-\nbt)}{(\Bt(\Omega^2+\Lambda^2)+\Bx\mathbf{q}^2+\tilde{\delta})^2}.
\eeq 

For the fermionic momenta $k_F$  does not scale under the RG\tcite{Polchinski1992}. $\ell$ scales towards the Fermi surface. The angle $\theta$ also does not scale.  The measure for fermionic type integrals is\tcite{Shankar1994}
\beq \label{fmeasure}
\int_k=k_F\int_{-\infty}^{\infty} \frac{d\omega}{2\pi} \int_{-\pi}^{\pi}\frac{d\theta}{2\pi}\int_{-k_{\rm UV}}^{k_{\rm UV}}\frac{ d\ell}{2\pi}=\frac{k_F\At\Lambda^2}{\Ax}\int^{\infty}_{-\infty}\frac{da}{2\pi}\int_{-\pi}^{\pi}\frac{d\theta}{2\pi}\int_{-Y}^{Y}\frac{db}{2\pi}.
\eeq
The frequency and momenta have been rescaled $a=\omega/\Lambda$, $b=\Ax\ell/\At\Lambda$. $Y$  is defined $Y=\Ax k_{\rm UV}/\At\Lambda$.

For integrals over purely bosonic frequencies and momenta, no prior knowledge of the Fermi surface structure is assumed thus both $q_x$ and $q_y$ must scale under the RG. The measure for bosonic type integrals is
\beq \label{bmeasure}
\int_q=\int_{-\infty}^{\infty} \frac{d\Omega}{2\pi} \int_{-\infty}^{\infty}\frac{dq_x}{2\pi}\int_{-\infty}^{\infty}\frac{ dq_y}{2\pi}=\frac{\At^2\Lambda^3}{\Ax^2}\int_{-\infty}^{\infty} \frac{da}{2\pi} \int_{-\infty}^{\infty}\frac{db_x}{2\pi}\int_{-\infty}^{\infty}\frac{ db_y}{2\pi}
\eeq
with the rescaled frequency and momenta $a=\Omega/\Lambda$, $b_x=\Ax q_x/\At\Lambda$, $b_y=\Ax q_y/\At\Lambda$. More precisely the absolute momentum of the boson should be constrained to $2k_F$ to remain in the particle-hole continuum, then the rescaled integral limits become $-\infty$ to $\infty$ only in the IR limit as $Y\rightarrow\infty$.

\section*{Flow equations with pure fermionic and bosonic contributions}
The flow equations for the bosonic parameters arise from diagrams with purely bosonic and fermionic lines. The flow of the boson mass term is given by 
\beq\label{massflow}
\sd\tilde{\delta}=-4(2+N_b)\tilde{\lambda}\Rd\int_q G_b(\Omega,\mathbf{q})+2\tilde{g}^2\Rd\int_k G_f^2(\omega,\mathbf{k}).
\eeq 
The flow of the boson self interaction is given by
\beq\label{intflow}
\sd\tilde{\lambda}=-4(8+N_b)\tilde{\lambda}^2\Rd\int_q G_b^2(\Omega,\mathbf{q})+\frac{\tilde{g}^4}{2}\Rd\int_k G_f^4(\omega,\mathbf{k}).
\eeq 
The flow of the bosonic frequency is given by
\beq \label{btflow}
\sd\Bt=\left.-2(2+N_b)\tilde{\lambda}\Rd\frac{\partial^2}{\partial\Omega'^2}\int_q G_b(\Omega+\Omega',\mathbf{q})+\tilde{g}^2\Rd\frac{\partial^2}{\partial\Omega'^2}\int_k G_f(\omega,\mathbf{k})G_f(\omega-\Omega',\mathbf{k})\right|_{\Omega'\rightarrow0}.
\eeq 
The flow of the bosonic momentum is given by
\beq \label{bxflow}
\sd\Bx=\left.-2(2+N_b)\tilde{\lambda}\Rd\frac{\partial^2}{\partial |\mathbf{q}'|^2}\int_q G_b(\Omega,\mathbf{q}+\mathbf{q}')+\tilde{g}^2\Rd\frac{\partial^2}{\partial|\mathbf{q}'|^2}\int_k G_f(\omega,\mathbf{k})G_f(\omega,\mathbf{k}-\mathbf{q}')\right|_{\mathbf{q}'\rightarrow0}.
\eeq 

The bosonic contribution to the flow of the boson mass is given by
\beq 
-4(2+N_b)\tilde{\lambda}\Rd\int_q G_b(\Omega,\mathbf{q})=-4(2+N_b)\tilde{\lambda}\int^{\infty}_{-\infty}\frac{d\Omega}{2\pi}\int_{-\infty}^{\infty}\frac{d^2\mathbf{q} }{(2\pi)^2}\frac{\Lambda\Bt(2-\nbt)}{(\Bt(\Omega^2+\Lambda^2)+\Bx\mathbf{q}^2+\tilde{\delta})^2}.
\eeq 
Rescaling to a dimensionless form the integral becomes
\beq 
-4(2+N_b)\tilde{\lambda}\Rd\int_q G_b(\Omega,\mathbf{q})=-8(2+N_b)\frac{\tilde{\lambda}\At^2}{\Bt\Ax^2}\int^{\infty}_{-\infty}\frac{da}{2\pi}\int_{0}^{\infty}\frac{db b }{2\pi}\frac{(2-\nbt)}{(1+a^2+\frac{c^2}{v^2}b^2+\delta)^2}.
\eeq 
$c^2=\Bx/\Bt$ and $v^2=\Ax^2/\At^2$. Performing the $a$ integration and making the substitution $u=\frac{c^2}{v^2}b^2$ 
\beq 
-4(2+N_b)\tilde{\lambda}\Rd\int_q G_b(\Omega,\mathbf{q})=-2(2+N_b)\frac{\tilde{\lambda}}{\Bx}\int_{0}^{\infty}\frac{du }{2\pi}\frac{(2-\nbt)}{(1+u+\delta)^{3/2}}=-(2+N_b)\frac{\tilde{\lambda}}{2\pi\Bx}\frac{(2-\nbt)}{\sqrt{1+\delta}}.
\eeq 
All further integrals containing only bosonic propagators are performed in the same fashion. It is therefore simple to compute higher order bosonic interactions such as sextic or higher order vertices.

The fermionic contribution to the flow of the boson mass is given by
\beq 
2\tilde{g}^2\Rd\int_k G_f^2(\omega,\mathbf{k})=4\tilde{g}^2k_F\int^{\infty}_{-\infty}\frac{d\omega}{2\pi}\int_{-\pi}^{\pi}\frac{d\theta}{2\pi}\int_{-k_{\rm UV}}^{k_{\rm UV}}\frac{d\ell}{2\pi}\chi'(\omega,\Lambda)\chi(\omega,\Lambda)\frac{i(2-\nft)\At\omega-(2-\nfx)\Ax\ell}{2(i\At\omega-\Ax\ell)^3}.
\eeq 
Performing the integral over $\theta$ and rescaling to a dimensionless form the integral becomes
\beq 
2\tilde{g}^2\Rd\int_k G_f^2(\omega,\mathbf{k})=\frac{4\tilde{g}^2k_F}{\Ax\At\Lambda}\int^{\infty}_{-\infty}\frac{da}{2\pi}\int_{-Y}^{Y}\frac{db}{2\pi}\frac{-2a^4}{(a^2+1)^3}\frac{i(2-\nft)a-(2-\nfx)b}{2(ia-b)^3}.
\eeq 
Now performing the integrals over $a$ and $b$ the result is
\beq 
2\tilde{g}^2\Rd\int_k G_f^2(\omega,\mathbf{k})=\frac{\tilde{g}^2k_F}{4\pi\Ax\At\Lambda}\left(\frac{(2-\nft)Y(1+4Y)}{(1+Y)^4}+\frac{3(2-\nfx)Y^3}{(1+Y)^4}\right).
\eeq 
Naively the integral above vanishes in the IR limit $Y\rightarrow\infty$; however, introducing the dimensionless parameter $N=k_F/k_{\rm UV}$ an additional factor of $Y$ is required to obtain the correct power of $\Lambda$:
\beq 
2\tilde{g}^2\Rd\int_k G_f^2(\omega,\mathbf{k})=\frac{\tilde{g}^2N}{4\pi\Ax^2}\left(\frac{(2-\nft)Y^2(1+4Y)}{(1+Y)^4}+\frac{3(2-\nfx)Y^4}{(1+Y)^4}\right)\overset{Y\rightarrow\infty}{=}\frac{3\tilde{g}^2N}{4\pi\Ax^2}(2-\nfx).
\eeq 

The flow of the dimensional mass (\ref{massflow}) is then
\beq 
\sd\tilde{\delta}=-(2+N_b)\frac{\tilde{\lambda}}{2\pi\Bx}\frac{(2-\nbt)}{\sqrt{1+\delta}}+\frac{3\tilde{g}^2N}{4\pi\Ax^2}(2-\nfx)
\eeq
Rescaling the parameters to the dimensionless forms (M10), and neglecting $\nfx$ on the right-hand side, gives the mass beta function (M11).

Considering the flow of the bosonic self interaction (\ref{intflow}) the fermionic contribution is of order $\mathcal{O}(Y^{-2})$ and vanishes in the IR. Calculating the bosonic contribution the dimensional flow equation is
\beq 
\sd\tilde{\lambda}=(8+N_b)\frac{\tilde{\lambda}^2}{4\pi\Bx\Bt\Lambda^2}\frac{(2-\nbt)}{(1+\delta)^{3/2}}
\eeq
leading to the beta function (M12).

Considering the fermionic contributions to flow equations (\ref{btflow}) and (\ref{bxflow}), the contribution to (\ref{bxflow}) vanishes due to the momentum derivatives causing the integral to be of order $\mathcal{O}(Y^{-2})$.  The bosonic contributions to the flow equations (\ref{btflow}) and (\ref{bxflow}) both vanish. Thus the anomalous dimension $\nbx=0$ throughout the flow. The surviving fermionic contribution of $\mathcal{O}(Y^{0})$ to (\ref{btflow}) leads to the anomalous dimension $\nbt=g^2(1-\nfx/2)$  in the $Y\rightarrow\infty$ limit.

\section*{Flow equations with mixed bosonic and fermionic lines}

The Yukawa vertex and fermion self energy terms are given by diagrams with mixed internal lines. The flow of the Yukawa vertex is given by
\beq \label{yukflow}
\sd\tilde{g}=-N_b\tilde{g}^3\Rd\int_kG_f^2(\omega,\mathbf{k})G_b(\omega,\mathbf{k}).
\eeq
The flow of the fermion frequency term is given by
\beq \label{atflow}
\sd\At=\left.N_b\tilde{g}^2\Rd\frac{\partial}{i\partial\omega'}\int_k G_f(\omega,\mathbf{k})G_b(\omega-\omega',\mathbf{k})\right|_{\omega'\rightarrow0}.
\eeq
The flow of the fermion momentum term is given by
\beq \label{axflow}
\sd\Ax=-\left.N_b\tilde{g}^2\Rd\frac{\partial}{\partial\ell'}\int_k G_f(\omega,\mathbf{k})G_b(\omega,\mathbf{k}-\mathbf{k}')\right|_{\mathbf{k}'\rightarrow k_F}.
\eeq
For diagrams with mixed internal lines the momentum transfers $\mathbf{q}$ and $\mathbf{k}-\mathbf{k}'$ should be indistinguishable in the bosonic lines. The momentum transfer is given by
\beq\label{ktransfer}
(\mathbf{k}-\mathbf{k}')^2=(\ell-\ell')^2+2k_F^2(1-\cos{\theta})+[2k_F(\ell+\ell')+2\ell\ell'](1-\cos{\theta}).   
\eeq

This suppresses the contribution to the flow equations for all values of theta other than $1-\cos{\theta}$ of the order $\mathcal{O}(1/k_F^2)$. Therefore the term in square brackets can be discarded as it is of order $\mathcal{O}(1/k_F)$\tcite{Fitzpatrick2013}.

This can be seen by considering the flow of the Yukawa vertex where $\ell'$ has been set to zero in (\ref{ktransfer}):
\begin{multline}
	\sd\tilde{g}=-N_b\tilde{g}^3k_F\int_{-\infty}^{\infty} \frac{d\omega}{2\pi} \int_{-\pi}^{\pi}\frac{d\theta}{2\pi}\int_{-k_{\rm UV}}^{k_{\rm UV}}\frac{ d\ell}{2\pi}\left[\frac{\chi^2(\omega,\Lambda)}{(i\At\omega-\Ax\ell)^2}\frac{\Lambda\Bt(2-\nbt)}{(\Bt(\Omega^2+\Lambda^2)+\Bx(\ell^2+2k_F^2(1-\cos{\theta}))+\tilde{\delta})^2} \right.\\\left.-\frac{\chi'(\omega,\Lambda)\chi(\omega,\Lambda)}{(i\At\omega-\Ax\ell)^2}\frac{(2-\nft)}{(\Bt(\Omega^2+\Lambda^2)+\Bx(\ell^2+2k_F^2(1-\cos{\theta}))+\tilde{\delta})}   \right].
\end{multline}
Rescaling $\omega$ and $\ell$ the flow equation becomes
\begin{multline}
	\sd\tilde{g}=-\frac{N_b\tilde{g}^3N Y}{\Ax^2\Bt\Lambda^2}\int_{-\infty}^{\infty} \frac{da}{2\pi} \int_{-\pi}^{\pi}\frac{d\theta}{2\pi}\int_{-Y}^{Y}\frac{ db}{2\pi}\left[\frac{2a^4}{(a^2+1)^3}\frac{1}{(b-ia)^2}\frac{(2-\nft)}{(1+a^2+\frac{c^2}{v^2}(b^2+2N^2Y^2(1-\cos{\theta}))+\delta)}\right.\\\left.+\frac{2a^4}{(a^2+1)^2}\frac{1}{(b-ia)^2}\frac{(2-\nbt)}{(1+a^2+\frac{c^2}{v^2}(b^2+2N^2Y^2(1-\cos{\theta}))+\delta)^2} \right].
\end{multline}
For $N$ large the $1-\cos{\theta}$ term contributes only when $\theta$ is small. Therefore taking the first term in the Taylor expansion for $1-\cos\theta\approx\theta^2/2$ and approximating the angular integral limits $\pm1/N$  the parallel momenta are constrained such that the scaling of the Yukawa coupling keeps the constant relevant term. Rescaling $b'=\theta NY$ the integral becomes
\begin{multline}
	\sd\tilde{g}=-\frac{N_b\tilde{g}^3}{\Ax^2\Bt\Lambda^2}\int_{-\infty}^{\infty} \frac{da}{2\pi} \int_{-Y}^{Y}\frac{db'}{2\pi}\int_{-Y}^{Y}\frac{ db}{2\pi}\left[\frac{2a^4}{(a^2+1)^3}\frac{1}{(b-ia)^2}\frac{(2-\nft)}{(1+a^2+\frac{c^2}{v^2}(b^2+b'^2)+\delta)}\right.\\\left.+\frac{2a^4}{(a^2+1)^2}\frac{1}{(b-ia)^2}\frac{(2-\nbt)}{(1+a^2+\frac{c^2}{v^2}(b^2+b'^2)+\delta)^2} \right].
\end{multline}
The limit $Y\rightarrow\infty$ can then be taken prior to performing the integrals. In the IR this integral is then indistinguishable from an integral over bosonic momenta $\mathbf{q}$ with the measure (\ref{bmeasure}). 

To compute the above integral it is simpler to consider the general integral
\beq 
\int_{-\infty}^{\infty} \frac{da}{2\pi} \int_{-\infty}^{\infty}\frac{db'}{2\pi}\int_{-\infty}^{\infty}\frac{ db}{2\pi}\frac{a^2}{(\beta a^2+\gamma)}\frac{1}{(b-ia)^2}\frac{1}{(1+a^2+z^2(b^2+b'^2)+\delta)}
\eeq
with $z^2=c^2/v^2$. Computing the $b'$ and $b$ integrals
\beq 
\int_{-\infty}^{\infty} \frac{da}{2\pi}\frac{a^2}{(\beta a^2+\gamma)}\left( \frac{z|a|}{4(1+(1-z^2)a^2+\delta)^{3/2}}- \frac{1}{2\pi(1+(1-z^2)a^2+\delta)}- \frac{za\arctan( \frac{za}{\sqrt{1+(1-z^2)a^2+\delta}})}{2\pi(1+(1-z^2)a^2+\delta)^{3/2}}\right).
\eeq
To evaluate the integral containing the arctangent overestimate the integral with the leading order term in the expansion $\arctan(x)\approx x$. For the flow equations (\ref{yukflow}-\ref{axflow}) the only surviving contribution when taking the necessary $\beta$, $\gamma$ or $\delta$ derivatives in the $z\rightarrow0$ limit corresponding to the overdamping of the boson is
\beq 
\int_{-\infty}^{\infty} \frac{da}{2\pi}\frac{a^2}{(\beta a^2+\gamma)}\frac{-1}{2\pi(1+(1-z^2)a^2+\delta)}=-\frac{1}{4\pi}\frac{1}{\beta\sqrt{1-z^2}\sqrt{1+\delta}+\sqrt{\beta\gamma}(1-z^2)}.
\eeq
Taking $z\rightarrow0$ and defining the function $F(\beta,\gamma,\delta)$ all the flow equations with mixed internal lines can be computed from
\beq 
F(\beta,\gamma,\delta)=-\frac{1}{4\pi}\frac{1}{\beta\sqrt{1+\delta}+\sqrt{\beta\gamma}}.
\eeq
The flow equation (\ref{yukflow}) becomes 
\beq 
\sd\tilde{g}=\left.-\frac{N_b\tilde{g}^3}{\Ax^2\Bt\Lambda^2}\left[(2-\nft)\pder{\beta}\pder{\gamma}+(2-\nbt)\pder{\beta}\pder{\delta}\right]F(\beta,\gamma,\delta)\right|_{\beta,\gamma\rightarrow1}. \label{flow1}
\eeq
The flow equation (\ref{atflow}) becomes 
\beq 
\sd\At=\left.\frac{N_b\tilde{g}^2\At}{\Ax^2\Bt\Lambda^2}\left[2(2-\nft)\pder{\beta}\pder{\delta}+2(2-\nbt)\pder{\gamma}\pder{\delta}+(2-\nbt)\frac{\partial^2}{\partial\delta^2}\right]F(\beta,\gamma,\delta)\right|_{\beta,\gamma\rightarrow1}. \label{flow2}
\eeq
The flow equation (\ref{axflow}) becomes 
\beq 
\sd\Ax=\left.\frac{N_b\tilde{g}^2}{\Ax\Bt\Lambda^2}\left[(2+\nfx)\pder{\gamma}+(2-\nbt)\frac{\partial^2}{\partial\delta^2}\right]F(\beta,\gamma,\delta)\right|_{\beta,\gamma\rightarrow1}. \label{flow3}
\eeq
Utilizing the definitions of the anomalous dimensions within the main paper, rescaling $\tilde{g}$, and neglecting the fermionic anomalous dimensions on the right-hand sides of (\ref{flow1}--\ref{flow3}) leads to the anomalous dimensions (M14), (M17) and (M16) respectively.

\end{widetext}

\end{document}